\documentclass{ws-mpla}
\begin{document}
\markboth{Abramovsky V.A.,Dmitriev A.V}{Hard non-vacuum reggeon in CKMT model}
\title{HARD NON-VACUUM REGGEON IN THE CKMT MODEL}
\author{\footnotesize V.A.ABRAMOVSKY, A.V. DMITRIEV}
\address{Novgorod State University, B. S.-Peterburgskaya Street 41,\\
Novgorod the Great, Russia,\\ 173003}
\maketitle

\begin{abstract}
The CKMT model describing the nucleon structure function $F_2(x,Q^2)$ in the framework of conventional Regge theory with smooth soft-hard pomeron is modified.
Smooth soft-hard non-vacuum reggeon dependence on incoming photon virtuality $Q^2$ is introduced. This dependence has the same functional form as the smooth
soft-hard functional $Q^2$ dependence for pomeron in original CKMT model. In regeon of low $W^2$ better agreement with experimental data is observed.
\end{abstract}

The CKMT model \cite{ckmt} uses Regge theory for description of the proton structure function $F_2(x,Q^2)$ in the region
of low $Q^2<5 GeV^2$. At higher values of $Q^2$ the corresponding Regge parametrization is used as initial condition for
DGLAP evolution.
 In this paper we review only low $Q^2$ region up to $5 GeV^2$, where perturbative corrections is still neglible.

The CKMT model \cite{ckmt} describes the proton structure function as the sum of
 vacuum and non-vacuum contributions:
\begin{equation}
F_2(x,Q^2) = F_v(x,Q^2) + F_{nv}(x,Q^2).
\label{eq:eq1}
\end{equation}
For the Pomeron contribution:
\begin{equation}
\begin{array}{l}
F_v(x,Q^2) = A\cdot x^{-\Delta(Q^2)}\cdot(1-x)^{n(Q^2)+4}
\cdot\left({Q^2\over Q^2+a}\right)^{1+\Delta(Q^2)}\\
n(Q^2) = {3\over2}\cdot\left(1+{ Q^2\over Q^2+c}\right)
\end{array}
\end{equation}
where the $x$$\rightarrow 0$
behavior is determined by an effective intercept
of the Pomeron,~$\Delta$,
which depends on $Q^2$. This dependence was
parametrized \cite{ckmt} as following:
\begin{equation}
\Delta (Q^2) = \Delta_0\cdot\left(1+{\Delta_1\cdot Q^2
\over Q^2+\Delta_2}\right).
\label{eq:eq3}
\end{equation}
Thus, for low values of $Q^2$, $\Delta$ is close
to the effective value found from analysis of hadronic total cross-sections
($\Delta=\Delta_0\sim0.08$), while for high values of $Q^2$,
$\Delta$ takes the hard Pomeron value,
$\Delta=\Delta_{bare}\sim0.2-0.25$. The
parametrization for the non-vacuum term corresponded to the secondary
reggeons ($f$, $A_2$) contribution is:
\begin{equation}
F_{nv}(x,Q^2) = B\cdot x^{-\Delta_R}\cdot(1-x)^{n(Q^2)}
\cdot\left({Q^2\over Q^2+b}\right)^{1+\Delta_R},
\label{eq:eq4}
\end{equation}
in original CKMT model \cite{ckmt}. Here the $x$$\rightarrow$0
behavior is determined by the secondary reggeon intercept
$\Delta_R$ estimated as $\Delta_R\sim-0.5.$.

 Equations (\ref{eq:eq1})-(\ref{eq:eq4}) describe pure CKMT model. We save all assumptions that were done above,
 but take into account the possible changing of non-vacuum reggeon parameters on incoming photon virtuality $Q^2$.
 Dependence of vertex function already had been included in (\ref{eq:eq4}), and can not be changed without breaking
 down commutation with real photon scattering.

So, the main idea of this paper is to introduce $Q^2$ dependence of non-vacuum reggeon intercept $\Delta_R$ in the
 same way, as in the case of Pomeron:
\begin{equation}
\Delta_R(Q^2) = \Delta_{R0}\cdot\left(1+{\Delta_{R1}\cdot Q^2
\over Q^2+\Delta_{R2}}\right).
\label{eq:eq4_5}
\end{equation}

We keep most parameters of the model the same as in the recent version \cite{ckmt1} CKMT model.
They are:$A=0.1301$,$a=0.2628$,$\Delta_0=0.09663$,$\Delta_1=1.9533$,$\Delta_2=1.1606$,$c=3.5489$,$b=0.384$.
Values of $B_u$ and $B_d$ were got, as in Ref.[\refcite{ckmt1}]:
\begin{center}
$B_u$=1.1555, $B_d$=0.1722.
\end{center}

So, we have only to vary parameters $\Delta_{R1}$ and $\Delta_{R2}$, because value
$\Delta_R$ at $Q^2=0$ is fixed well from real photon cross-section to value $\Delta_{R0}=-0.585$.

\begin{figure}[th]
\centerline{\psfig{file=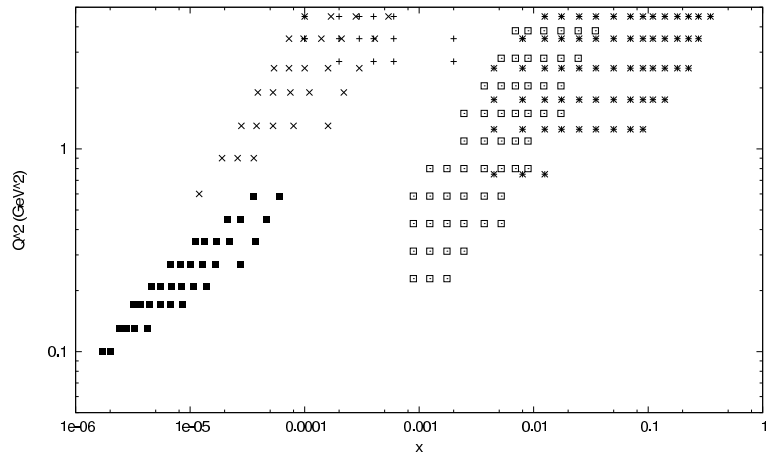,width=4.5in}}
\vspace*{8pt}
\caption{Kinematical region, used in the fitting procedure. Data used: BPC95 (blacks squares), E665 (white squares),NMC (*),
SVX (x), ZEUS94 (+).}
\label{fig:fig1}
\end{figure}

Fitting was done in the kinematical region, shown in Fig. \ref{fig:fig1} (data set was got from Ref.[\refcite{durpdg}]).
This region covers all available values of $x$ and covers $Q^2$ up to values, where perturbative corrections must be done.

After fitting we have derived next values:
\begin{equation}
\Delta_{R1}=0.188;\Delta_{R2}=1 GeV^2
\label{eq:eqres}
\end{equation}

\begin{figure}[th]
\centerline{\psfig{file=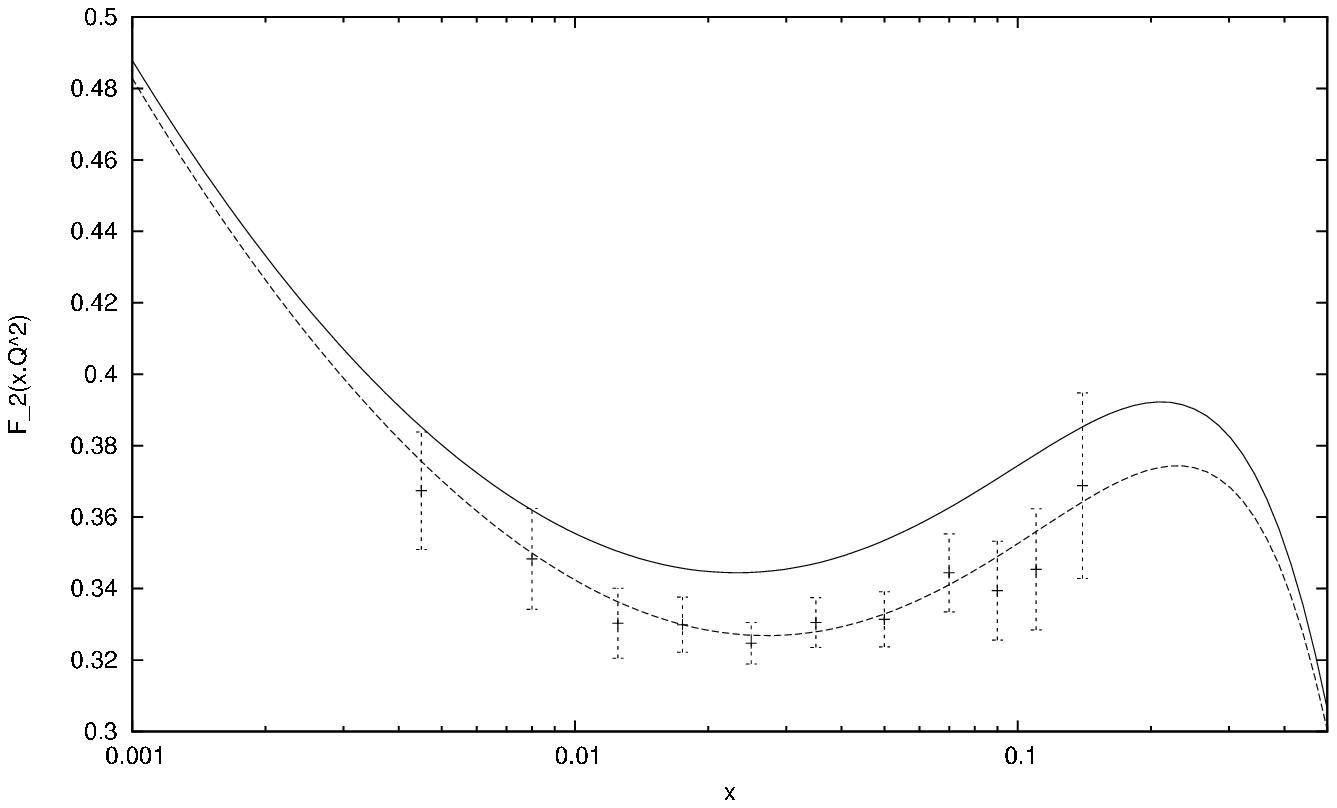,width=4in}}
\centerline{\psfig{file=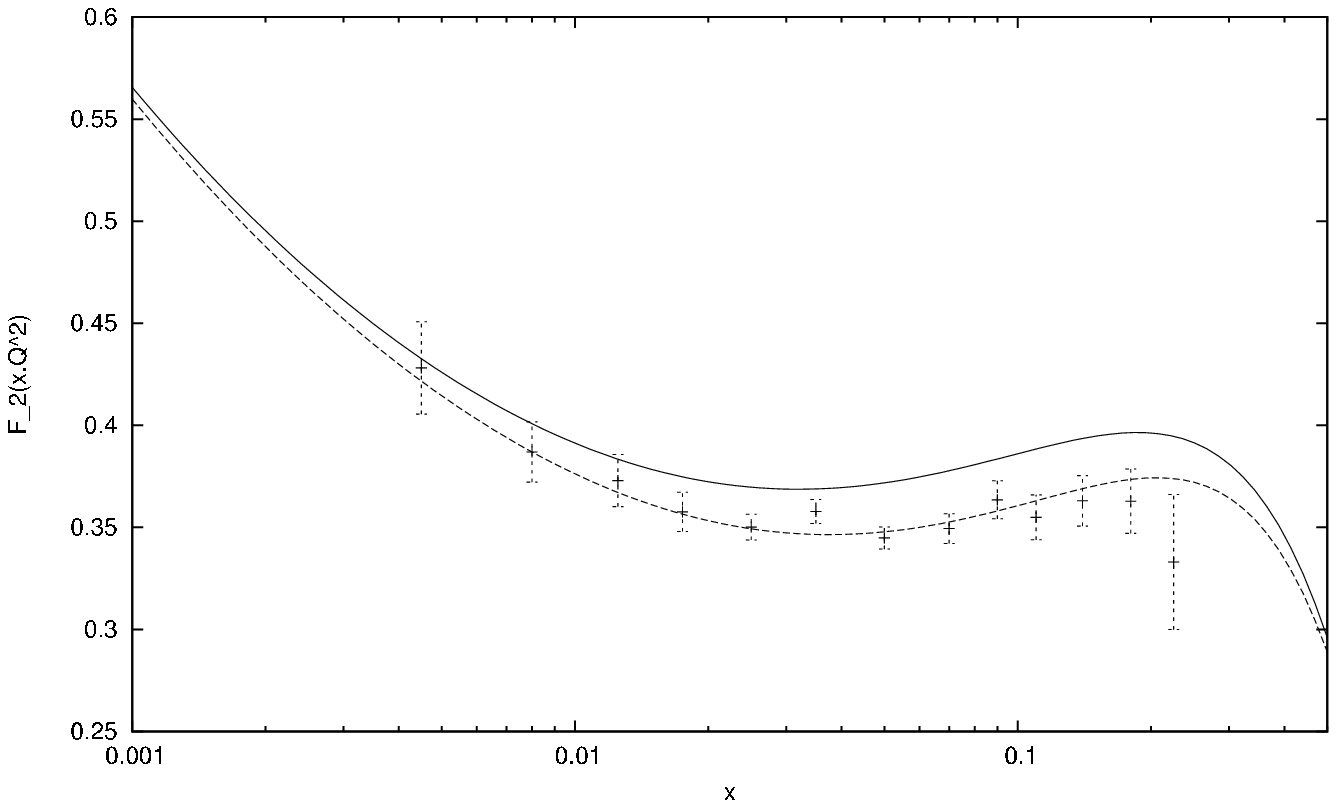,width=4in}}
\vspace*{8pt}
\caption{$F_2$($x$,$Q^2$) vs $x$ for $Q^2=1.75 Gev^2$(top) and for $Q^2=2.5 Gev^2$(bottom). Theoretical curves have been
obtained with the original CKMT model (full line) and with our model (dashed line).}
\label{fig:fig2a}
\end{figure}

\begin{figure}[th]
\centerline{\psfig{file=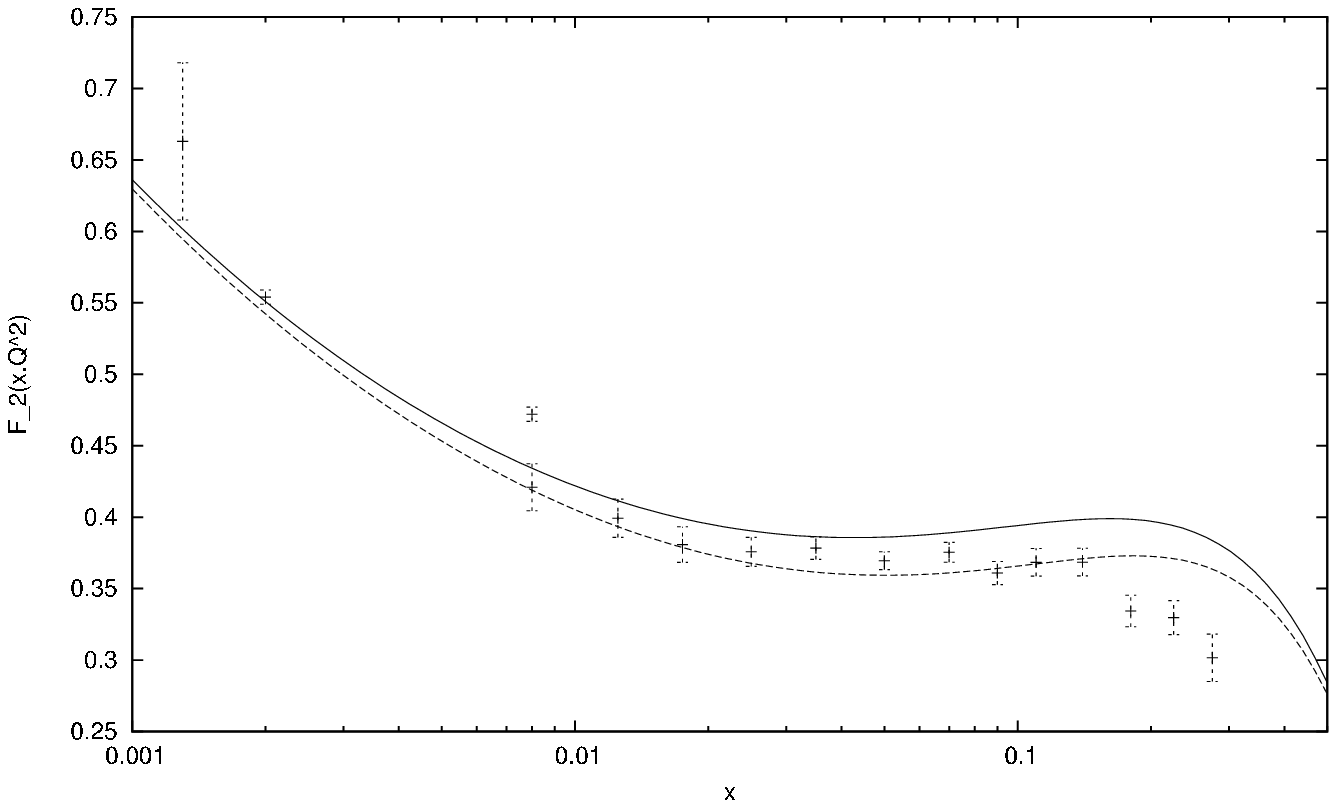,width=4in}}
\centerline{\psfig{file=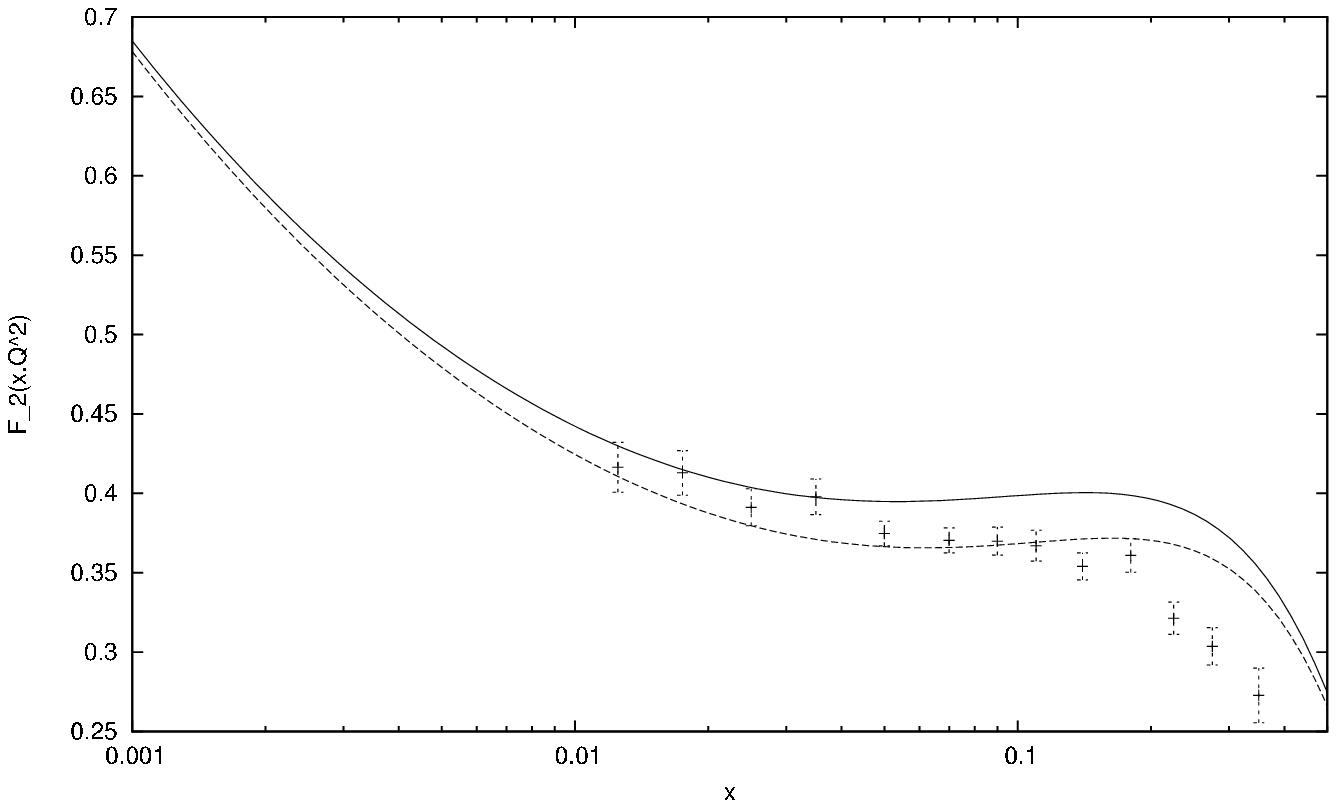,width=4in}}
\vspace*{8pt}
\caption {$F_2$($x$,$Q^2$) vs $x$ for $Q^2=3.5 Gev^2$(top) and for $Q^2=4.5 Gev^2$(bottom). Theoretical curves have been
obtained with the original CKMT model (full line) and with our model (dashed line).}
\label{fig:fig2c}
\end{figure}

At moderate values of $Q^2$ our approach gives predictions, which is different from pure CKMT model.
It can be seen from the Figs. \ref{fig:fig2a}-\ref{fig:fig2c}, that our approach gives significantly better agreement in
 the region of moderate $x$(or low $W^2$). Moreover, test $\chi^2$ became significantly better on this data set, for pure
  CKMT $\chi^2/n.d.f=5.5$ and for our model $\chi^2/n.d.f=3.4$.

\begin{figure}[th]
\centerline{\psfig{file=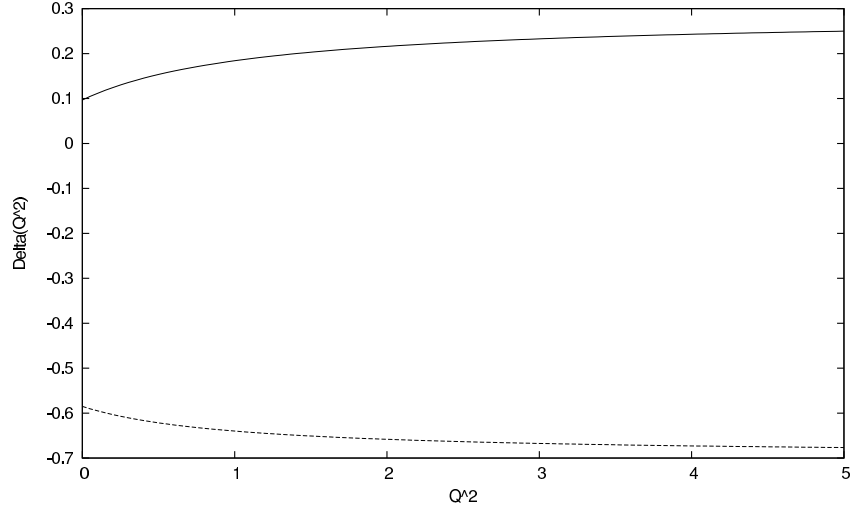,width=5in}}
\vspace*{8pt}
\caption {$\Delta(Q^2)$ vs $Q^2$ for pomeron (solid) and for non-vacuum reggeon(dashed).}
\label{fig:fig2aa}
\end{figure}

It is necessary to stress that the absolute values of vacuum trajectory shift and non-vacuum one

\begin{equation}
\begin{array}{l}
\delta\Delta_P \equiv \Delta_P (Q^2=\infty ) -  \Delta_P (Q^2=0) = \Delta_0 \cdot \Delta_1 \sim 0.188 \\
\delta\Delta_R \equiv \Delta_R (Q^2=\infty ) -  \Delta_R (Q^2=0) = \Delta_{R0} \cdot \Delta_{R1} \sim -0.109
\end{array}
\end{equation}
are the same order of magnitude. It can be cleary seen from Fig. \ref{fig:fig2aa}.

Let`s discuss possible physical motivations of changing intercept $\Delta_R$ with changing $Q^2$.
In the CKMT model \cite{ckmt1} the decreasing of the vacuum trajectory at low virtualities $Q^2$ is explained by the
contribution of the pomeron cuts. This explanation is not valid for nonvacuum trajectory. We can estimate this effect
 in simple eikonal model. In this model we suppose that vertex functions don`t depend on pomeron transverse momenta.

 At high virtuality $Q^2$ the contributions of pomeron cuts are suppressed. So the non-vacuum reggeon contribution to  proton
structure function is:
\begin{equation}
\begin{array}{c}
F_{nv}(x,Q^2)=\frac{1}{W^2}A_{Rbare}(Y,\overline q=0)=g_R^2e^{\Delta_{Rbare} Y}\\
Y\equiv ln(W^2),
\end{array}
\end{equation}
In our notation (\ref{eq:eq4_5})
\begin{equation}
\Delta_{Rbare}=\Delta_{R0}(1+\Delta_{R1})
\end{equation}
To estimate this amplitude at $Q^2=0$ we have to account the pomeron cuts. In impact parameter representation we have
\begin{equation}
\begin{array}{l}
\frac{1}{W^2}A(Y,\overline b)=\frac{2(2\pi)^2g_R^2}{4\alpha_R^{\prime}Y}e^{\Delta_{Rbare}Y-\frac{b^2}{4\alpha_R^{\prime}Y}}
\cdot e^{-\frac{g^2}{4\alpha^{\prime}Y}e^{\Delta Y-\frac{b^2}{4\alpha^{\prime}Y}}}
\end{array}
\end{equation}
In the black disk limit last term is
\begin{equation}
e^{-\frac{g^2}{4\alpha^{\prime}Y}e^{\Delta Y-\frac{b^2}{4\alpha^{\prime}Y}}}=\left\{
\begin{array}{ll}
0 & \textrm{if } b^2<4\Delta\alpha^{\prime} Y^2\\
1 & \textrm{if } b^2>4\Delta\alpha^{\prime} Y^2
\end{array}\right.
\end{equation}
and we get for $A(Y,\overline q)$ at optical point:
\begin{equation}
\frac{1}{W^2}A(Y,\overline q=0)=g_R^2e^{\Delta_{Rbare} Y}e^{-\Delta Y}\\
\label{eq:eq5}
\end{equation}
Comparing (\ref{eq:eq5}) with the effective non-vacuum reggeon contribution at $Q^2=0$
\begin{equation}
\frac{1}{W^2}A(Y,\overline q=0)=g_R^2e^{\Delta_{R0} Y}
\end{equation}
we have
\begin{equation}
\Delta_{R0}=\Delta_{Rbare}-\Delta,\Delta_{R0}<\Delta_{Rbare}
\end{equation}
and we expect $\Delta_{R1}<0$ in this model.
This is in disagreement with our result (\ref{eq:eqres}), so we must
reject this eikonal based  explanation.

The possible explanation is that at high $Q^2$ we have perturbative fermion ladder with intercept $\Delta_{Rbare}=-1$, and
 non-perturbative effects shift interception to values $\Delta_{R0} \sim -0.5$, so
\begin{equation}
\Delta_{R0}>\Delta_{Rbare}.
\label{eq:eq6}
\end{equation}
In our notation, this model predicts  $\Delta_{R1}>0$, which is in qualitative agreement with our result (\ref{eq:eqres}).

Strict theoretical explanation of the trajectory shift is a serious problem deviated from the framework of this paper.

In conclusion we stress that $Q^2$ dependence of the non-vacuum reggeon intercept gives the better agreement
with experimental data. So this effect leads to the different initial conditions for DGLAP evolution than pure CKMT model
and may be important in definition of structure function $F_2(x,Q^2)$ at high $Q^2$.

\section*{Acknowledgments}
We thank N.Prikhod`ko for useful discussions.
This work was supported by RFBR Grant RFBR-03-02-16157a and grant of Ministry for Education E02-3.1-282

\end{document}